\documentclass[aps,prl,groupedaddress,superscriptaddress,twocolumn,longbibliography]{revtex4-2}
\usepackage{graphicx}% Include figure files
\graphicspath{{../figures/}}
\usepackage{dcolumn}% Align table columns on decimal point
\usepackage{bm}% bold math
\usepackage{hyperref}% add hypertext capabilities
\usepackage{orcidlink}
\usepackage{siunitx}
\usepackage{amsmath}
	\usepackage[normalem]{ulem} %makes underlining possible: \uline \uwave \sout
	\definecolor{mgreen}{RGB}{1,123,0}

\def \mm{\mathrm{m}}
\def \mum{\mu\mathrm{m} }

\def \mHz{\mathrm{Hz}}

\def \mnK{\mathrm{nK}}

\def \mms{\mathrm{ms}}

\def \mum{\mu \mathrm{m}}
\def \mref{\mathrm{ref}}

\def \mm{\mathrm{m}}

\def \mHz{\mathrm{Hz}}

\def \mnK{\mathrm{nK}}

\def \mms{\mathrm{ms}}

\def \mm{\mathrm{m}}
\def \mum{\mu\mathrm{m} }

\def \mHz{\mathrm{Hz}}

\def \mnK{\mathrm{nK}}

\def \mms{\mathrm{ms}}

\newcommand{\corauthor}[2]{
    \author{#1}
    \email{#2}
}

\begin{document}

\preprint{APS/123-QED}

\title{Observation of Shapiro steps in an ultracold atomic Josephson junction}

\author{Erik Bernhart\orcidlink{0000-0002-2268-0540}}
\affiliation{Department of Physics and Research Center OPTIMAS, Rheinland-Pf\"alzische Technische Universit\"at Kaiserslautern-Landau, 67663 Kaiserslautern, Germany}

\author{Marvin R\"ohrle\orcidlink{0000-0002-2257-1504}}
\affiliation{Department of Physics and Research Center OPTIMAS, Rheinland-Pf\"alzische Technische Universit\"at Kaiserslautern-Landau, 67663 Kaiserslautern, Germany}

\author{Vijay Pal Singh\orcidlink{0000-0001-8881-978X}}
\affiliation{Quantum Research Centre, Technology Innovation Institute, Abu Dhabi, UAE}

\author{Ludwig Mathey}
\affiliation{Zentrum f\"ur Optische Quantentechnologien and Institut f\"ur  Quantenphysik, Universit\"at Hamburg, 22761 Hamburg, Germany}
\affiliation{The Hamburg Centre for Ultrafast Imaging, Luruper Chaussee 149, Hamburg 22761, Germany}
\author{Luigi Amico}%{luigi.amico@tii.ae}
\affiliation{Quantum Research Centre, Technology Innovation Institute, Abu Dhabi, UAE}
\affiliation{Dipartimento di Fisica e Astronomia, Universit\`a di Catania, Via S. Sofia 64, 95123 Catania, Italy}
\affiliation{INFN-Sezione di Catania, Via S. Sofia 64, 95127 Catania, Italy}

\corauthor{Herwig Ott\orcidlink{0000-0002-3155-2719}}{ott@physik.uni-kl.de}
\affiliation{Department of Physics and Research Center OPTIMAS, Rheinland-Pf\"alzische Technische Universit\"at Kaiserslautern-Landau, 67663 Kaiserslautern, Germany}

\date{\today}% It is always \today, today,

\begin{abstract}
  
The current-voltage characteristic of a driven superconducting Josephson junction displays discrete steps. This phenomenon, called the Shapiro steps, forms today's voltage standard! Here, we report the observation of Shapiro steps in a driven Josephson junction in a gas of ultracold atoms. We demonstrate that the steps exhibit universal features, and provide insight into the microscopic dissipative dynamics that we directly observe in the experiment. We find that the steps are directly connected to phonon emission and nucleation of solitonic excitations, whose dynamics we follow in space and time. 
The experimental results are underpinned by extensive numerical simulations based on classical-field dynamics and may enable metrological and fundamental advances. %represent the transfer of the voltage standard to the realm of ultracold quantum gases.  

\end{abstract}
\maketitle

%\tableofcontents

%%%%%%%%%%%%%%%%%%%%%%%%%%%%%%%%
%%%%%%%%%%%%%%%%%%%%%%%%%%%%%%%%
% Define all figures as commands. This allows for faster rearranging.

\newcommand{\figoverview}{
\begin{figure*}
	\includegraphics[width=2\columnwidth]{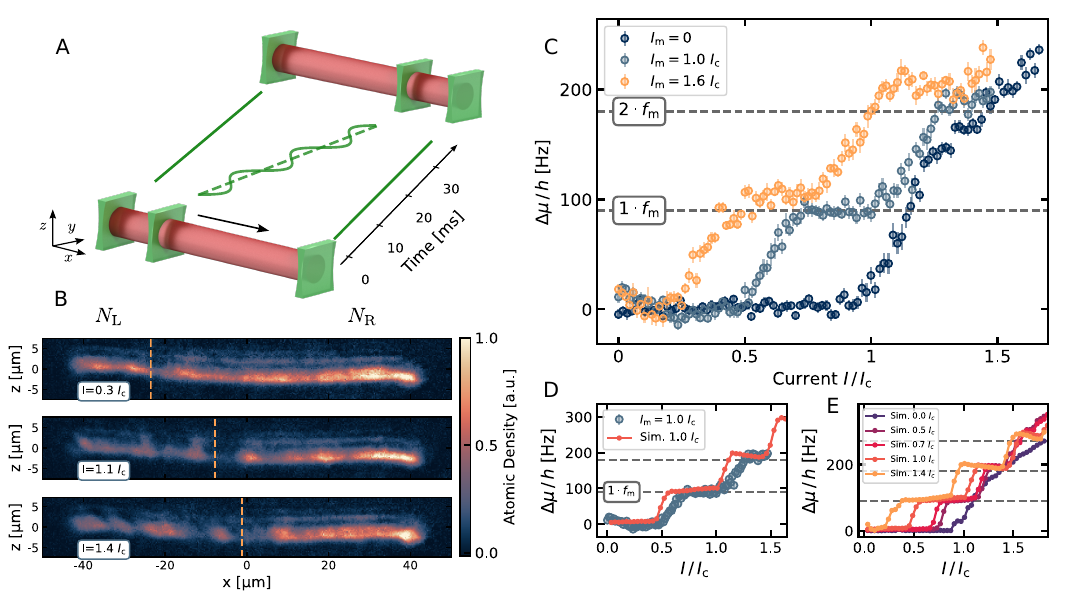}
	\caption{
	\textbf{Shapiro steps in an ultracold atomic Josephson junction.} (\textbf{A}) The system consists of a cylindrically shaped superfluid which is split by a movable optical barrier creating a weak link. The barrier as well as the two confining end-caps are realized with tightly focused laser beams. In the measurement protocol, the barrier is moved linearly (dc particle current) through the superfluid with an additional harmonic modulation (ac particle current), which is sketched in the figure. (\textbf{B}) At the end of the protocol, a real space absorption image of the atomic cloud is taken, using  matter wave imaging  \cite{Asteria_2021,supp}. From the atom number imbalance between the two superfluids, their chemical potential difference is derived. The pictures are taken with a modulation current $I_\mathrm{m} = \SI{0.8}{} \, I_\mathrm{c}$. The position of the barrier is marked as a dashed orange line. (\textbf{C}) Measured Shapiro steps for $f_\mathrm{m} = \SI{90}{Hz}$, and different modulation amplitudes $I_\mathrm{m}$. The horizontal lines indicate the ideal Shapiro step height, given by Eq.\,\ref{eq:deltamu}. Every data point is the average of about $25$ measurement runs, the vertical bars indicate the error of the mean.  (\textbf{D}) Comparison between experiment and classical-field simulation at $T=\SI{35}{nK}$ \cite{supp}. (\textbf{E}) Simulations of the system at different $I_\mathrm{m}$ \cite{supp}.}
	\label{fig:overview}
\end{figure*}
}

\newcommand{\figfrequencydep}{
\begin{figure}
	\includegraphics[width=\columnwidth]{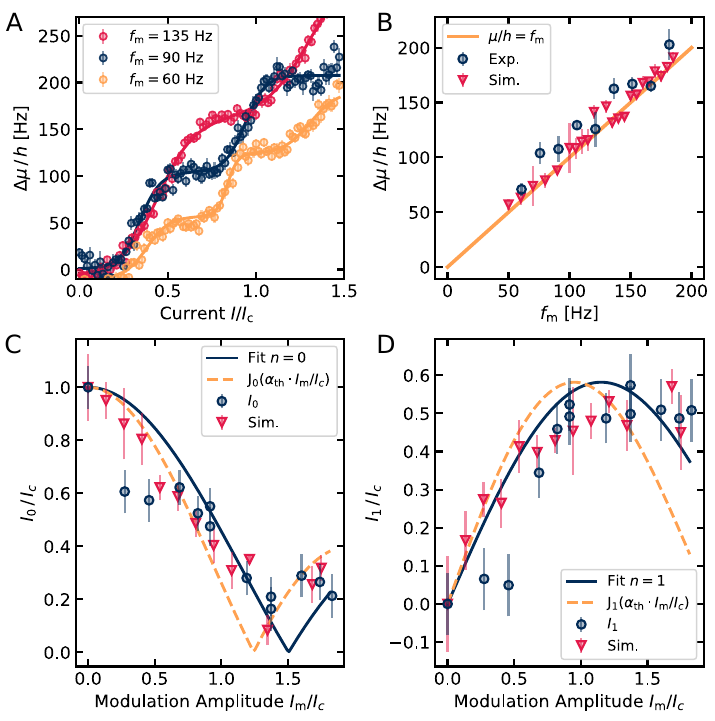}
	\caption{
    \textbf{Characteristics of the Shapiro steps.}
	(\textbf{A}) Dependence of the step height on the modulation frequency $f_\mathrm{m}$, where the step height is determined using sigmoid fits (continuous lines). %For better visibility, part of the data points are rendered transparent.
    (\textbf{B}) Measurements of the step height (dots) are compared to the numerical simulations (triangles) and the prediction $\Delta \mu \, / \, h = f_\mathrm{m}$ (continuous line). 
	(\textbf{C} and \textbf{D}) Measurements (dots) and simulation (triangles) of zeroth (\textbf{C}) and first (\textbf{D}) step widths ($I_0$ and $I_1$) for varying $I_\mathrm{m}/I_c$. 
      The solid blue lines are a fit to the experimental data, following Eq. \ref{eq:besselfunc}, where both curves are fitted simultaneously with the same fit parameter $\alpha_\mathrm{fit}$, yielding $\alpha_\mathrm{fit}=2.15\pm0.08$. The dashed orange lines indicate the theoretical prediction, following Eq. \ref{eq:besselfunc}, see text.
    }
	\label{fig:freqdep}
\end{figure}
}

\newcommand{\microscopic}{
\begin{figure*}
	\includegraphics[width=1.6\columnwidth]{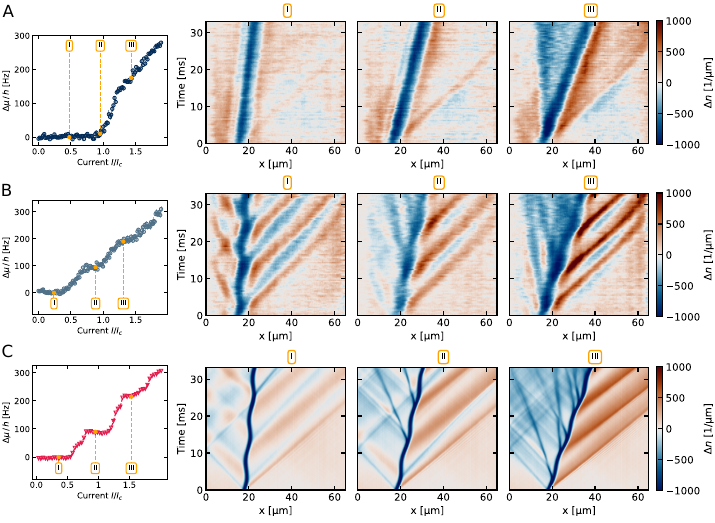}
	\caption{
    \textbf{Microscopic system dynamics.}
    Time evolution of the atomic density in the experiment (A, B) and in the simulation (C). (\textbf{A}) Constant barrier velocity. We show the time evolution for three different currents, indicated in the $I-\Delta \mu$ plot on the left. (\textbf{B}) Constant plus periodically modulated barrier velocity (Shapiro protocol), where $I_\mathrm{m} = \SI{0.8}{}I_\mathrm{c}$ and $f_\mathrm{m} = \SI{90}{Hz}$. (\textbf{C}) Classical-field simulations of the driven junction, with $I_\mathrm{m} = \SI{0.8}{}I_\mathrm{c}$ and $f_\mathrm{m} = \SI{90}{Hz}$.  Each horizontal line in the images corresponds to a transversely integrated absorption image, which are stacked together. The color code indicates the relative change of the line density with respect to a reference image without barrier.
	}
	\label{fig:microscopic}
\end{figure*}

\begin{figure*}
\includegraphics[width=2\columnwidth]{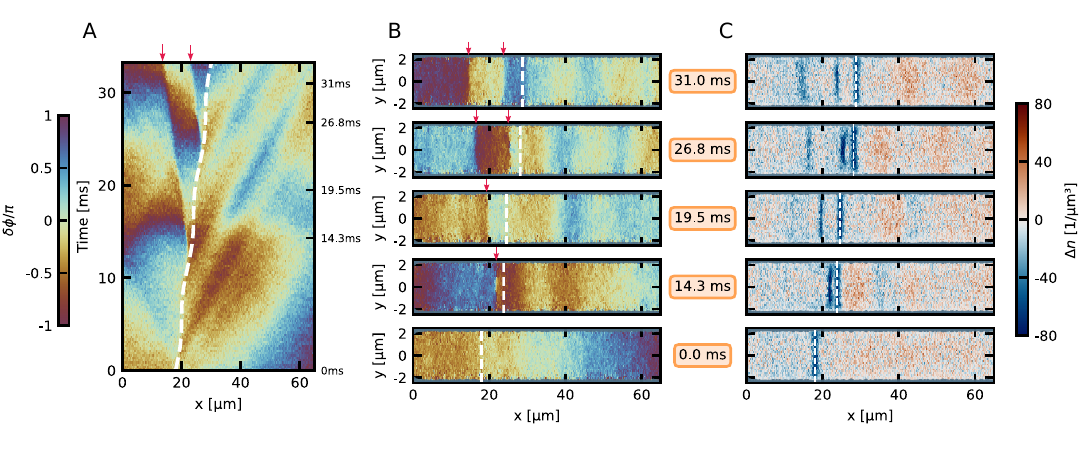}
\caption{
\textbf{Dynamics of solitonic excitations.}
(\textbf{A}) Time evolution of the phase $\delta \phi (x, t) =  \phi (x, t) - \phi_\mathrm{m}( t)$ along the central $x$ line of a single trajectory of the system for $I/I_c= 0.9$, $I_\mathrm{m}/I_c= 0.8$ and $f_\mathrm{m}=90\, \mathrm{Hz}$, where $\phi_\mathrm{m}( t)$ is obtained after averaging the wavefunction over $x$. 
The periodic barrier motion (white dashed line) modifies the local phase near the barrier, which results in the creation of solitonic excitations moving to the left. 
The red arrows in (A) and (B) are a guide to the eye for the position of the solitonic excitations.   
(\textbf{B}) Snapshots of the phase profile $\delta \phi (x,y)$ of a single trajectory and (\textbf{C}) the averaged density profile $\Delta n(x, y)$ for the central $z$ plane at times $t=0$, $14.3$, $19.5$, $26.8$, and $31\, \mms$. During each driving cycle the barrier motion (white vertical dashed line) emits one solitonic excitation, whose subsequent time evolution manifests itself as density dips in (C). The positions of the phase jumps in (A) are marked with red arrows. At time $14.3$\,ms, the phase profile of a vortex ring is visible \cite{supp}. There are also phononic excitations that propagate in both forward and backward directions. 
In (C) we show the relative change of the density with respect to a reference cloud without barrier. In (B, C), the thermal wings outside the Thomas-Fermi region of the condensate are shaded in gray.
 }
\label{fig:soliton}
\end{figure*}
}


%\section{Introduction}

The Josephson effect is one of the most fundamental phenomena in quantum science and technology, %transport effects. 
%originally discovered for superconductors, the  effect 
featuring a dissipation-less electrical current through a tunnel barrier between two superconductors by virtue of the phase difference of the traversing Cooper pairs' wave function.  
Above a critical value of the current, a finite voltage  occurs across the junction as a result of the formation of superconducting quasi-particles \cite{JOSEPHSON1962251}.  
The Josephson effect and Josephson junctions have played a prominent role in understanding the notion of macroscopic quantum coherence, which led to important  technological applications such as superconducting quantum interference devices (SQUIDs) \cite{tinkham2004introduction,10.1063/1.2354545, RevModPhys.71.631,Barone,leggett1987macroscopic} they are also the core ingredients of superconducting qubits \cite{Clarke2008,PhysRevA.76.042319,Johnson2011,Arute2019}. 
If an additional microwave field is applied across the Josephson junction, a staircase current-voltage characteristic emerges, referred to as  Shapiro steps \cite{ShapiroOriginal}.
The origin of this effect is photon-assisted tunneling of Cooper pairs through the junction \cite{Barone}. 
Because the height of the steps depends only on the frequency of the microwave radiation, Planck's constant, and the electric charge,  Shapiro steps are nowadays used as the voltage standard - see e.g. \cite{popel1992josephson}. Shapiro steps in superfluid $^3$He have been investigated in Ref.\,\cite{Shapiro_superfluid}.

The Josephson effect can also be observed in ultracold quantum gases \cite{Oberthaler_2005}.   
This enables the fabrication of atomtronic circuits, which have developed into their own field of research in recent years \cite{Atom_squid,roadmap_atomtronic, Colloquium_Atomtronics}. 
Additionally, the high degree of control and flexibility over ultracold quantum gases provides unprecedented access to the microscopic physics underlying the system. In particular, Josephson physics has been recently implemented and extensively studied in ultracold atomic systems, both theoretically \cite{Giovanazzi_2000,Meier_Zwerger_2001,Singh_2020,PhysRevA.104.023316,Grond_2011,Kohler_2003} as well as experimentally in bosonic \cite{Oberthaler_2005, LeBlanc_2011,Campbell_2016,Ji_2022} and fermionic \cite{Roati_dc_jj_scince_2015,Moritz_jj_science} systems. 
In implementations with ultracold atoms, the Josephson current-phase relation is realized by moving the barrier through an ultracold gas which is otherwise at rest \cite{Roati_acdc_jj_2020, Roati_acdc_jj_2021}.
Shapiro steps in ultracold gases are predicted to appear if the linear translation of the barrier (corresponding to a dc particle current) is combined with a periodic modulation of the position of the barrier (corresponding to an ac particle current), thus emulating the external microwave radiation of the superconducting Josephson junction \cite{singh2023shapiro}.

\figoverview

Here, we demonstrate the emergence of Shapiro steps at an atomic Josephson junction and quantify their global and local properties. 
The steps occur in the chemical potential difference, in conjunction with a density imbalance across the junction. 
We confirm that the step heights in the chemical potential difference $\Delta \mu$ are quantized by the external driving frequency $f_\mathrm{m}$, i.e., $\Delta \mu = nh f_\mathrm{m}$, where $n$ is the step index and $h$ is Planck's constant.
By spatially resolving the atomic density, we study the microscopic  dynamics and observe the propagation of phonons and density depletions emerging from the barrier, which we identify as solitonic excitations.
This paves the way for studying the microscopic dissipative dynamics of Shapiro steps and other related Josephson transport effects.
Furthermore, given that the magnitude of the step height relates the chemical potential difference to a tunable frequency, and that the density imbalance across the barrier is directly measurable, our results suggest a protocol to measure the equation of state of strongly correlated superfluids.% thus advancing the study of superfluids with a novel approach.

\section{Experimental System}

The experimental setup and the Shapiro protocol are sketched in Fig.\,\ref{fig:overview}. 
We prepare a $^{87}\mathrm{Rb}$ Bose-Einstein condensate in an elongated optical dipole trap. 
The condensate has a tube-like cylindrical geometry with harmonic transverse confinement, see Fig. \ref{fig:overview} A. 
The chemical potential is $\mu = h\times 1900\,s^{-1}$.  
A movable repulsive barrier with height $V_B = 0.45 \times \mu$ separates the condensate into two parts and thus realizes a weak link \cite{supp}.
We probe the system by absorption imaging. To achieve a high accuracy, we perform a matter wave imaging scheme \cite{Asteria_2021} in order to reduce the optical density \cite{supp}.

We implement the dc and ac driving protocol following the proposal in Ref.\,\cite{singh2023shapiro}. 
To this end, we move the barrier with constant velocity $v$ and an additional periodic modulation, 
\begin{equation}
	x(t) = v t + x_\mathrm{m} \sin(2 \pi f_\mathrm{m} t)\, ,
\end{equation}
where $x_\mathrm{m}$ is the modulation amplitude, and $f_\mathrm{m}$ the modulation frequency.
We fix the driving time to $\SI{33}{ms}$, which means that the final position of the barrier depends on $v$.
At the end, we measure the atom number imbalance $z = (N_\mathrm{R} - N_\mathrm{L})/N$, 
where $N$ is the total atom number and $N_\mathrm{L}$ ($N_\mathrm{R}$) is the atom number on the left (right) hand side of the barrier. 
As a reference $z_{\mathrm{ref}}$, we also measure the corresponding imbalance without the barrier. 
Using a simulation of the Gross-Pitaevskii equation we map $\Delta z = z -z_{\mathrm{ref}}$ to the chemical potential difference $\Delta \mu$ \cite{supp}.
The barrier motion induces an atomic current relative to the barrier of value $I$, with $I = v I_\mathrm{c}/v_\mathrm{c}$, where $I_\mathrm{c}$ is the critical current and $v_\mathrm{c}$ is the critical velocity \cite{supp}.
The external current then reads $I_{\mathrm{ext}} = I + I_\mathrm{ac} = I +  I_\mathrm{m} \sin(2 \pi f_\mathrm{m} t)$, with the bias current $I$ and the modulation current $I_\mathrm{m}$.
The parameters that we vary in our protocol are $I$, $I_\mm$ and $f_\mm$.
In Fig. \ref{fig:overview} C we show the measurements of the current-chemical potential characteristics 
that are obtained without and with ac driving.  
From the undriven response we characterize the critical velocity and the critical current of the Josephson junction \cite{supp}. 
For the periodically driven case, the system features plateaus of constant chemical potential difference occurring at integer multiple of the driving frequency; the steps occur at lower $I$ for stronger $I_\mm$. 
We model our system with extensive classical-field simulations \cite{supp}, which capture the features of our driven Josephson junction well, see Fig.\,\ref{fig:overview}, D and E.

\figfrequencydep

In a superconducting Josephson junction where the Shapiro steps arise from photon-assisted tunneling of the Cooper pairs \cite{DayemPhotonAssTunel}, the voltage $V_n$ of the $n$-th Shapiro step is directly given by the energy quantization, i.e.,  
$V_n = n \Phi_0 f_\mathrm{m} = n f_\mathrm{m} h/(2e)$, 
where $\Phi_0$ is the flux quantum and $e$ the elementary charge \cite{Barone,PhysRevB.5.912}.  
For our neutral atom implementation this results in the quantization condition
\begin{equation}
\Delta \mu = nh f_\mathrm{m} \, .
\label{eq:deltamu}
\end{equation}
To confirm this fundamental relation we vary $f_\mm$ and determine the height $\Delta \mu$ of the first step, see Fig.\,\ref{fig:freqdep} A. In Fig.\,\ref{fig:freqdep} B we show the results for the simulation and for the experiment, by averaging over several realization with different modulation amplitudes. 
The simulation reproduces the relation trend accurately, but the experiment measures a $10\%$ larger slope. The reason for this small deviation lies in the uncertainties of the experimental parameters (atom number and trapping potential) which enter the calculation of $\mu$ (see also \cite{supp}). Taking this into account, we conclude that we indeed confirm the behavior $\Delta \mu= h f_\mm$ (Fig.\,\ref{fig:freqdep} B).
%From an application point of view, 
Our results show that the driving protocol is capable of 
producing a predetermined chemical potential difference, which may enable the transfer of the voltage standard to the realm of ultracold quantum gases. \cite{Kohler_2003, HamiltonVoltStandard}.
In addition, our results are useful in situations where the relation between the atomic density and the chemical potential, $\mu (n)$, is not known. 
Measuring the density difference at both sides of the junction directly reveals $d \mu (n)/ {dn}$, which is referred to as the inverse thermodynamic density of states. 
Performing stepwise measurements by decreasing the density allows for the determination of the equation of state $\mu (n)$. 

Next, we analyze the width of the Shapiro steps by varying the modulation amplitude $I_\mathrm{m}/I_\mathrm{c}$ at a fixed modulation frequency of $f_\mathrm{m} = \SI{90}{Hz}$ \cite{supp}. 
In Fig. \ref{fig:freqdep}, C and D, the widths of the zeroth and first steps display a Bessel-function behavior as a function of $I_\mathrm{m}/I_\mathrm{c}$, which is in agreement with the results of numerical simulations \cite{supp}. 
To support this observation we use the analytical prediction of an ac voltage driven Josephson junction, 
which yields the step width $I_n = I_c |J_n(V_\mm/V_n)|$, 
where $J_n$ is the Bessel function of the $n$-th order and $V_\mm$ the modulation voltage \cite{Barone,PhysRevLett.26.426}.
For an ac current driven junction this can be mathematically transformed into
\begin{equation}
\label{eq:besselfunc}
I_n =  I_\mathrm{c} \left| J_n \left(\alpha \frac{ I_\mathrm{m} }{I_\mathrm{c}}\right) \right| \, ,
\end{equation}
with $\alpha=R I_\mathrm{c}/f_\mathrm{m}$.
We use the independently determined system parameters $R$ and $I_\mathrm{c}$, see \cite{supp}, to calculate this parameter, obtaining $\alpha_{\mathrm{th}}= 1.93 \pm 0.07$. 
Using $\alpha_\mathrm{th}$, we plot the result of Eq. \ref{eq:besselfunc} as the orange dashed line in Fig. \ref{fig:freqdep}, C and D. 
Notably, this fit parameter free model supports the overall trend of our measurements and simulations.

 %such that there is no free fitting parameter \cite{supp}.
%In Fig. \ref{fig:freqdep} C, D the results of Eq. \ref{eq:besselfunc} support the overall trend of our measurements and simulations. 

\microscopic

\section{Microscopic Dynamics}

In comparison to solid-state physics, the time and length scales in quantum gas experiments are orders of magnitude larger, enabling in situ study of collective excitations, quasi-particles and related microscopic phenomena.
To that end, we follow the time evolution of the real space atomic density.
Using only a constant barrier velocity, i.e., no ac particle current, we measure the dc and ac Josephson regime, see Fig.\,\ref{fig:microscopic} A.
Below the critical current no imbalance is created and only a single phononic density wave is emitted from the initial acceleration of the barrier (I and II).
The group velocity of the emitted phonons coincides with a reference measurement, where a single perturbation creates a phonon wave packet \cite{supp}.
In the ac Josephson regime above the critical current, atoms are pushed away by the barrier and a density depletion remains in its wake (III).

In the Shapiro regime instead, we see that the phase dynamics is characterized by two distinct collective excitations: phonons and localized density depletions (see Fig.\,\ref{fig:microscopic} B and C). 
Phonons are emitted at distinct times in both directions thanks to the oscillatory motion of the barrier. This is particularly visible on the zeroth plateau (I).
Increasing the constant current $I$ beyond the zeroth plateau yields a chemical potential difference and atom number imbalance; this phenomenon is characterized by the formation of 'depletion waves' that move backwards compared to the barrier motion (blue lines propagating to the left in II and III). 
These excitations are observed to have a smaller group velocity, i.e., a steeper slope \cite{supp}. By additionally using the phase information of the simulation in Fig.\,\ref{fig:soliton} A, we observe that such density depletions are accompanied by phase jumps across the density minimum and identify them as solitonic excitations \cite{Ku_2014}. 
Analyzing the transverse phase profile right after the emission \cite{supp}, we find that the solitonic exctiations manifest themselves as vortex rings.
%This is also supported experimentally by an analysis of the soliton velocity and the resulting density depletion in the soliton center, see supplementary material.
These solitonic excitations are responsible for most of the reduction in density, ultimately  causing the particle imbalance across the junction- see Fig. \ref{fig:soliton} B and C. 
Specifically, the simulation shows that for the $n$-th plateau $n$ solitonic excitations are on-average emitted backwards during one driving cycle, see panels II and III of Fig.\,\ref{fig:microscopic} C and \cite{supp}. This creates the reduced density in the wake of the moving barrier, and therefore gives rise to the resistive regime.
This statement is supported by the fact that the rate of emitted solitonic excitations indeed displays a step-like behavior in close parallel with  the staircase displayed by the chemical potential, see Fig. S13.  
Depending on the system geometry the emerging solitonic excitations can be different. For instance, in the theoretical proposal \cite{singh2023shapiro}, a larger 2D system was used and vortex-antivortex pairs are generated at the barrier.
The appearance of vortex rings have been seen in Ref.\,\cite{Roati_dc_jj_2018_phaseslips,Xhani_2020}, as well as the emission of solitons from a barrier in Ref.\,\cite{Engels_2007,Hakim_1997}.

We note that Shapiro steps have also been observed in a complementary experiment with ultracold fermions in the unitary regime and in a molecular BEC \cite{DelPace2024}. 
Whereas in our work, vortex rings are the relevant solitonic excitations in the microscopic dynamics, the different geometry in Ref.\,\cite{DelPace2024}\ leads to the formation of vortex-antivortex pairs. 

It is also instructive to compare the microscopic physical mechanism to that of superconductors. 
Our results effectively demonstrate that both  superconducting and atomic junctions are characterized by coherent dynamics leading to a synchronization of the phase difference with the external drive \cite{tinkham2004introduction}. \\
Whereas in superconductors, the tunneling of Cooper pairs in the voltage state is mediated by photons from the coherent driving field, the atomic superfluid couples to phonons when crossing the barrier at finite chemical potential difference. \\
As for the resistive channel, superconductors are characterized by an incoherent tunneling of quasi-particles (namely broken Cooper pairs), which are present in the superconductor at finite temperature \cite{Barone}. 
In contrast, in atomic superfluids, collective excitations such as solitons and vortices are the dominant dissipation mechanism, opening the resistive channel. 
As they can be observed directly in the experiment, their role for the microscopic physics of Shapiro steps can be studied under clean conditions. \\
In superconductors, collective excitations have been investigated in complementary studies on Josephson physics, in which phase slips \cite{Sivakov2003,Shaikhaidarov2022}, Josephson vortices \cite{Roditchev2015} and solitons \cite{Rajasekaran2016} have been observed experimentally. 
All these observations have been made with static or pinned excitations; direct observation of the dynamics of such excitations remains challenging.
However, we expect that collective phase-excitations might also play a role in the Shapiro physics of superconductors, especially in regimes, e.g. at very low temperatures, where the quasi-particle density is small.

\section{Discussion and Outlook}

We have experimentally demonstrated the emergence of Shapiro steps in an ultracold quantum gas by emulating the external microwave field and an alternating particle current across the barrier with a periodic modulation of the barrier position. 
Using cold atom techniques, we could monitor the coherent dynamics of the system with unprecedented accuracy. 
%This way, our work  opens up  new directions both in basic and applied physical science. 

Extensions of the current work include, different dimensions, geometries, and particle statistics, as well as superfluid mixtures, dipolar superfluids, and superfluids in optical lattices.
These can generate other types of solitons, such as bright solitons, or vortices, or defects related to an underlying lattice or competing orders.   
Indeed,  Eq.\,\ref{eq:deltamu} provides the possibility to measure the equation of state of strongly correlated many-body systems.
In the context of  superconducting Josephson junctions, our work motivates expanding the understanding of the microscopic dynamics of the Shapiro effect, including the role of collective excitations such as solitons and vortices in regimes where the quasi-particle density is low. Furthermore, our results demonstrate the potential of atomic superfluids to act as quantum simulators for superconducting circuits.

%On the application side, we remark that our system transfers the voltage standard to the field of ultracold quantum gases.
Towards the development of atomtronic technology, Shapiro steps can be used to fine tune quantum transport,  as they combine both superfluid and resistive transport.
Stacks of Shapiro steps can be used to create even larger differences in chemical potential over a whole system, or as a source for predetermined chemical potential differences.

\section{Acknowledgments}

We thank Giacomo Roati and Giulia Del Pace for useful discussions.
\textbf{Funding:}
We gratefully acknowledge financial support by the DFG within the SFB OSCAR (project number 277625399).
L.M. acknowledges support by the Deutsche Forschungsgemeinschaft (DFG, German Research Foundation), namely the Cluster of Excellence ‘Advanced Imaging of Matter’ (EXC 2056), Project No. 390715994. 
The project is co-financed by ERDF of the European Union and by ’Fonds of the Hamburg Ministry of Science, Research, Equalities and Districts (BWFGB)’.
\textbf{Author contributions:}
E.B., M.R. and H.O. conceived the study. E.B. and M.R. performed the experiment and analyzed the data. 
E.B. performed the GPE simulations and V.P.S. modeled classical-field simulations. 
H.O. supervised the experiment. L.M. and L.A. supervised the theoretical part of the project.
E.B. prepared the initial version of the manuscript. 
All authors contributed to the data interpretation and the writing of the manuscript.
\textbf{Competing interests:}
The authors declare no competing interests. 
\textbf{Data and materials availability:}
The study involved no new materials preparation. 
All data presented in this paper and simulation codes are available at \url{https://doi.org/10.26204/data/14} \cite{data_repo}.

%%%%%%%%%%%%%%%%%%%%%%%%%%%%%%%%%%%% 
%bibliography

\providecommand{\noopsort}[1]{}\providecommand{\singleletter}[1]{#1}%
%

%%%%%%%%%%%%%%%%%%%%%%%%%%%%%%%%%%%%%%%%%%%%%%%%%%%%%%%%%%%%%%%%%%%%%%%%%%%%%%%%%%%%%%%%%%%%%%%%%%%%%%%%%%%%%%%%%%%%%
%%%%%%%%%%%%%%%%%%%%%%%%%%%%%%%%%%%%%%%%%%%%%%%%%%%%%%%%%%%%%%%%%%%%%%%%%%%%%%%%%%%%%%%%%%%%%%%%%%%%%%%%%%%%%%%%%%%%%
%%%%%%%%%  Supplementary material

%\renewcommand{\bibsection}{\section*{Supplementary References}}

\newcommand{\beginsupplement}{
    \setcounter{table}{0}
    \renewcommand{\tablename}{Supplementary Table}
    \setcounter{figure}{0}
    \renewcommand{\figurename}{Figure S}
}
\clearpage
\beginsupplement

\section{Supplementary Materials}

\newcommand{\figstshbarrheight}{
\begin{figure*}
	\includegraphics[width=2\columnwidth]{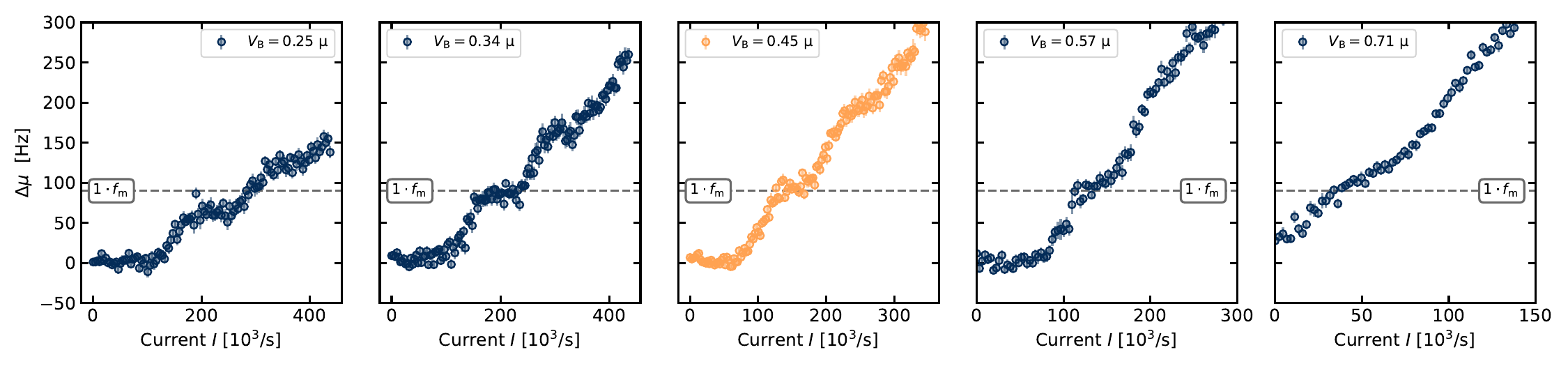}
	\caption{
	\textbf{Shapiro steps for different barrier heights.} Barrier heights between $0.25 \, \mu$ and $0.7 \mu$ for a modulation frequency of $f_\mathrm{m} = \SI{90}{Hz}$. The steps are clearly visible over the complete parameter range. However, deviations start to be visible at the edges of the parameter range (see text). 
	}
	\label{fig:stsh_barrier}
\end{figure*}
}

\newcommand{\figmeasresistance}{
\begin{figure}
	\includegraphics[width=0.7\columnwidth]{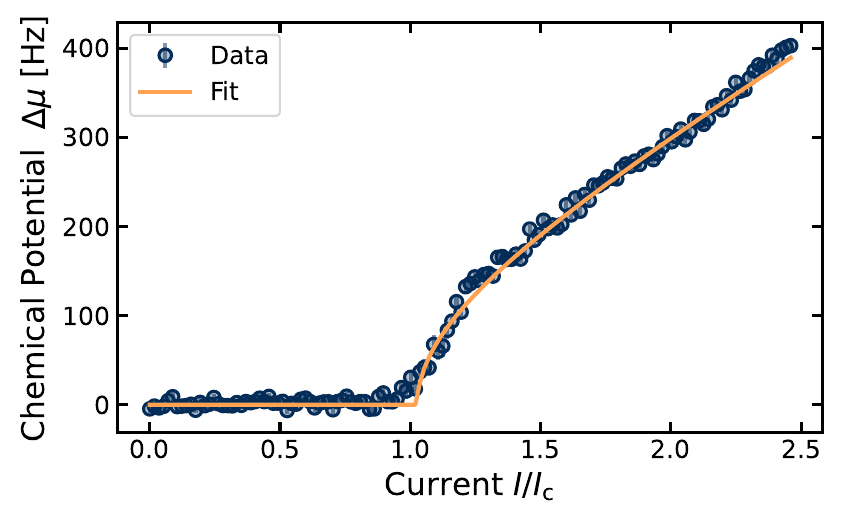}
	\caption{
    \textbf{Determination of the resistance $R$.}
	Critical velocity measurement with fit to get the resistance $R=\SI{{907 \pm 4}e-6}{} \times h $. 
	}
	\label{fig:measresistance}
\end{figure}
}

\newcommand{\figsystemcharc}{
\begin{figure}
	\includegraphics[width=1\columnwidth]{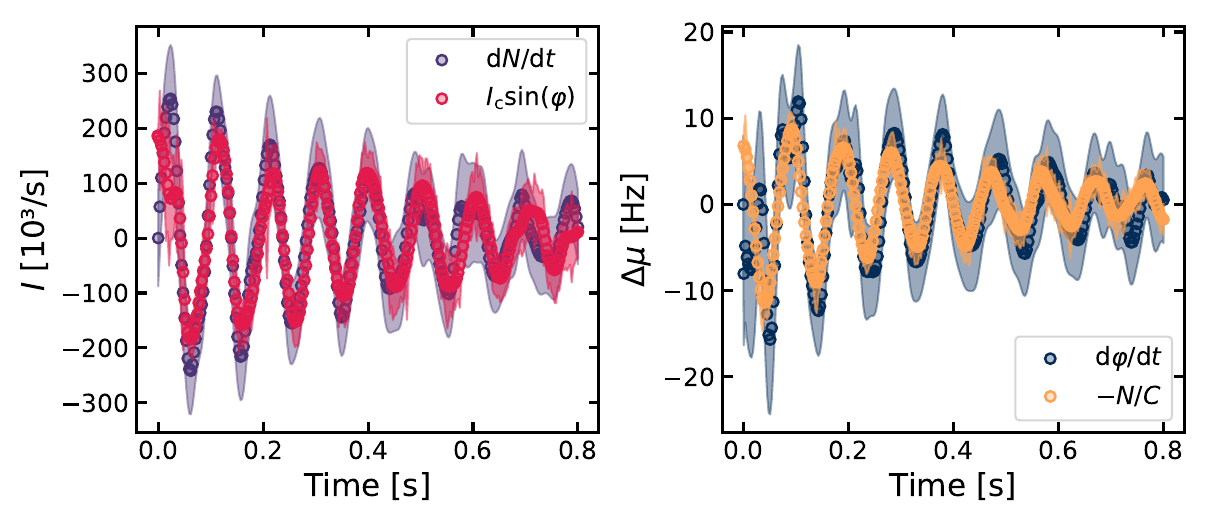}
	\caption{
	\textbf{Determination of the system parameters $I_c$ and $C$.} 
	We measure the atom number difference $N$ and the relative phase $\phi$ of a Josephson plasma oscillation.  
	Numerically differentiating $N$ and $\phi$ and subsequently fitting $I_c$ and $C$ to the equations $\frac{dN}{dt} = I_c \sin(\phi(t))$, $\frac{d\phi}{dt} = -\frac{N}{C}$, gives to $I_c = \SI{{192\pm 7}e3}{1/s}$ and $C = \SI{65.0 \pm 3.9}{s} \times 1/h$. 
	}
	\label{fig:systemcharac}
\end{figure}
}

\newcommand{\chempot}{
\begin{figure}
	\includegraphics[width=0.7\columnwidth]{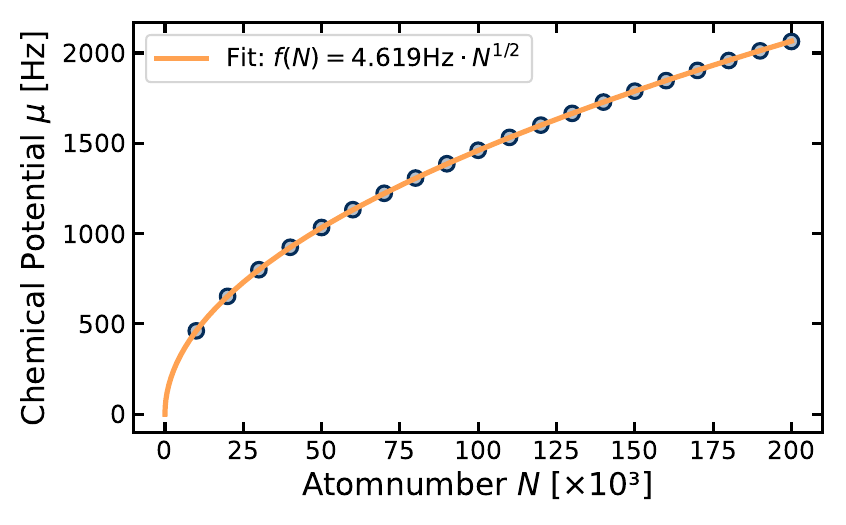}
	\caption{
	\textbf{Chemical potential $\mu$ in dependence of the total atom number.} The calculation was done using imaginary time evolution of the Gross-Pitaevskii equation for the given system parameters. 
	}
	\label{fig:chem_pot}
\end{figure}
}

\newcommand{\chempotcalib}{
\begin{figure}
	\includegraphics[width=0.7\columnwidth]{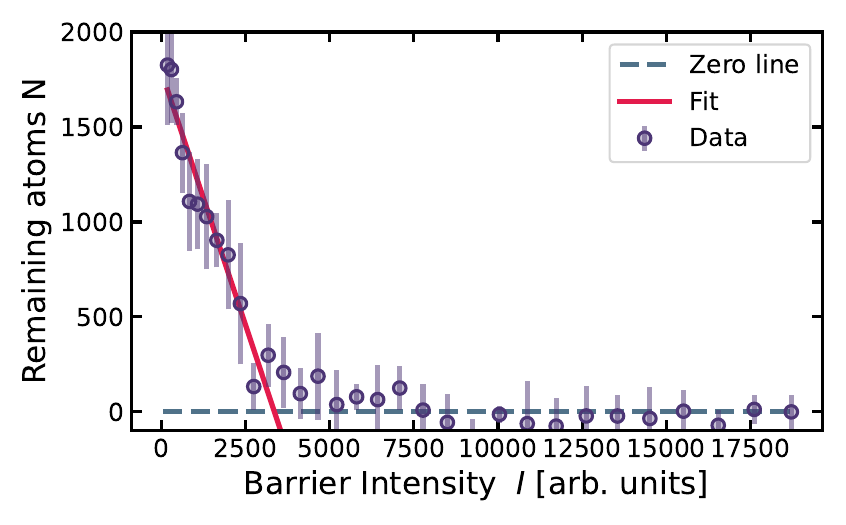}
	\caption{
	\textbf{Calibration of the barrier height.} A barrier block of $5 \times 8 $ spots is projected into the center of the BEC. The atom number in the block area is measured and plotted against the barrier intensity. The solid line is a linear fit, according to $n=\mu/g$. The intersection with the abscissa defines the light intensity, for which $V_0=\mu$  
	}
	\label{fig:chem_pot_calib}
\end{figure}
}

\newcommand{\rcsj}{
\begin{figure}
	\includegraphics[width=0.5\columnwidth]{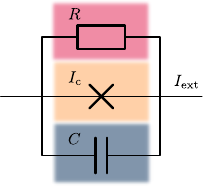}
	\caption{
	\textbf{RCSJ circuit.} RCSJ model circuit to describe a Josephson junction.
	}
	\label{fig:rcsj}
\end{figure}
}

\newcommand{\speedofsound}{
\begin{figure}
	\includegraphics[width=\columnwidth]{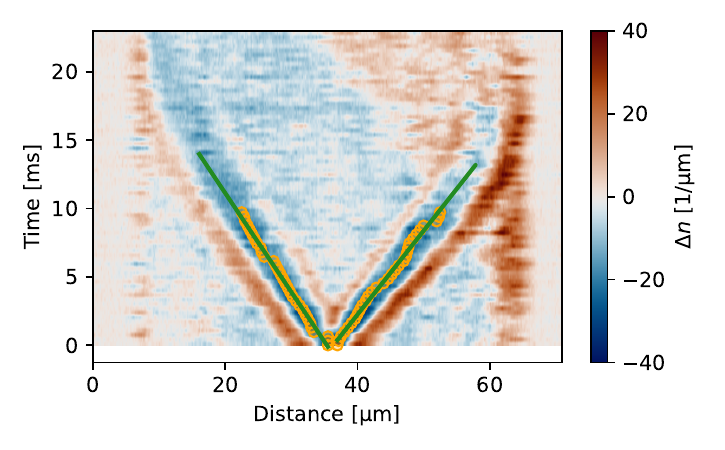}
	\caption{
	\textbf{Speed of sound measurement.} Ramp the barrier up to excite a density wave moving through the condensate and take absorption images to extract the velocity. For the wave of the right side $c_s = \SI{1.63 \pm 0.03 }{mm/s}$ and on the left side $c_s = \SI{-1.38 \pm 0.02 }{mm/s}$ 
	}
	\label{fig:speedofsound}
\end{figure}
}

\newcommand{\currentphase}{
\begin{figure}
	\includegraphics[width=0.7\columnwidth]{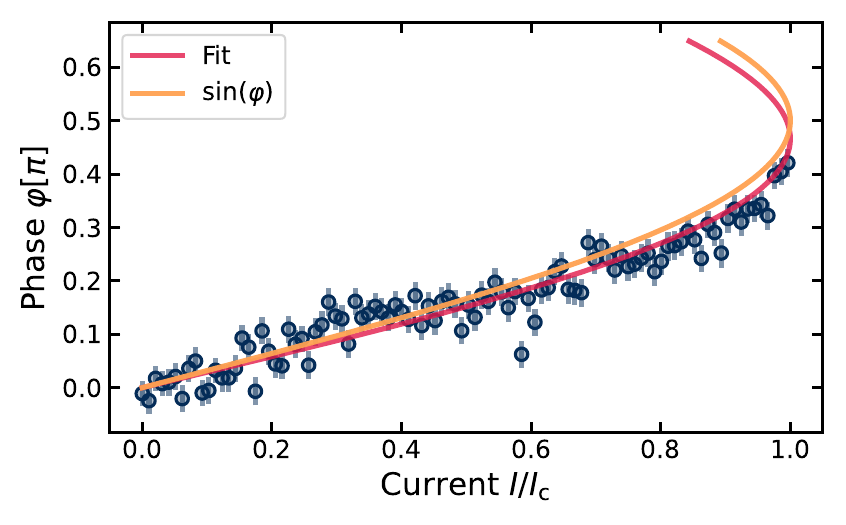}
	\caption{
 \textbf{Current-phase relation of the Josephson junction.}
	Measurement of the current-phase relation of the Josephson junction using the dc Josephson effect. The phase difference across the barrier is plotted against the dc current. The yellow curve represents the ideal current phase-relation $I=I_c\sin(\varphi)$. The red curve is a fit including the second order term (see text).  
	}
	\label{fig:cpr}
\end{figure}
}

\newcommand{\figstepwidthsingle}{
\begin{figure}
	\includegraphics[width=0.6\columnwidth]{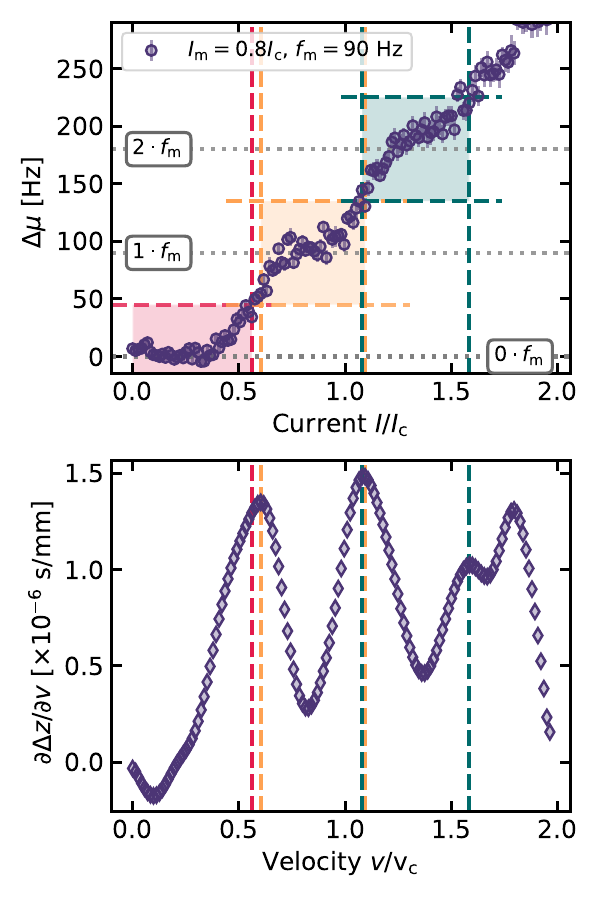}
	\caption{
	\textbf{Illustration of the step width evaluation} \textbf{Upper:} Shapiro Step dataset where the shaded areas mark $\Delta \mu = n f_\mathrm{m} \pm \epsilon$ , with $\epsilon = \SI{45}{Hz}$. \textbf{Lower:} First derivative of the Shapiro Step dataset, vertical lines mark the maxima of the derivative $D_\mathrm{M}$ in the area $\Delta \mu = n f_\mathrm{m} \pm \epsilon$. 
	}
	\label{fig:stepwidth_single}
\end{figure}
}

\newcommand{\figsimfreq}{
\begin{figure}
\includegraphics[width=1.0\columnwidth]{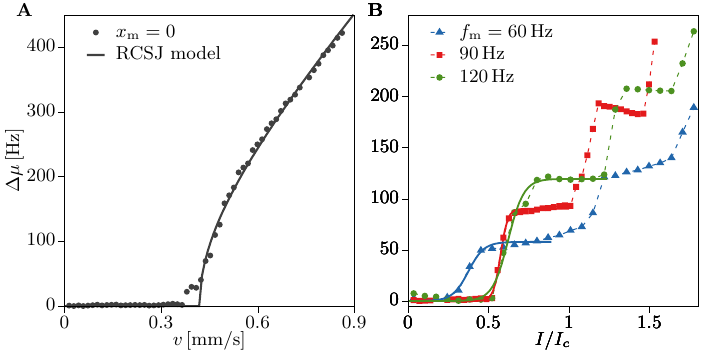}
\caption{
\textbf{Simulation without and with ac driving.} 
(\textbf{A}) Simulated response after $33\, \mms$ as a function of the dc velocity $v$, which we fit with the result of the RCSJ circuit model: $\Delta \mu  = R \sqrt{v^2 - v_c^2}$ using $R$ and $v_c$ as the fitting parameters. 
(\textbf{B}) Driven response as a function of the dc current $I/I_c$ for the driving frequency $f_\mm= 60$, $90$ and $120\, \mHz$ and the driving amplitude $I_\mm/I_c=0.9$. The heights of the first step are determined using sigmoid fits (continuous lines).  
	}
	\label{fig:simfreq}
\end{figure}
}

\newcommand{\figsimsoliton}{
\begin{figure}
\includegraphics[width=1.0\columnwidth]{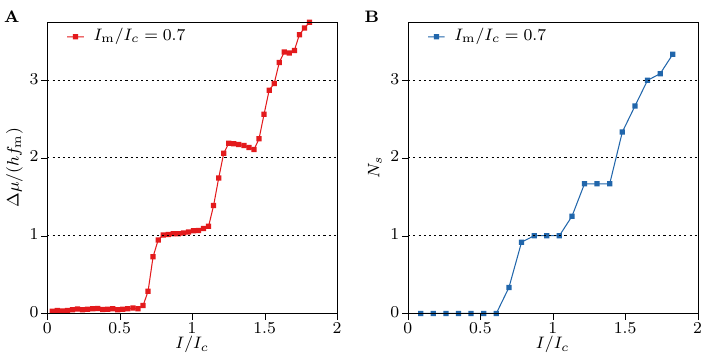}
\caption{
\textbf{Shapiro step mechanism.} 
(\textbf{A}) Simulation result of the current-chemical potential characteristic for $f_\mm= 90\, \mHz$ and $I_\mm/I_c=0.7$. 
(\textbf{B}) Total number of solitonic excitations, $N_s$, created in the system for the same parameters as in A, which is determined after averaging over a few samples. 
	}
	\label{fig:simsoliton}
\end{figure}
}

\newcommand{\figsolitonvelocity
}{
\begin{figure}
\includegraphics[width=1.0\columnwidth]{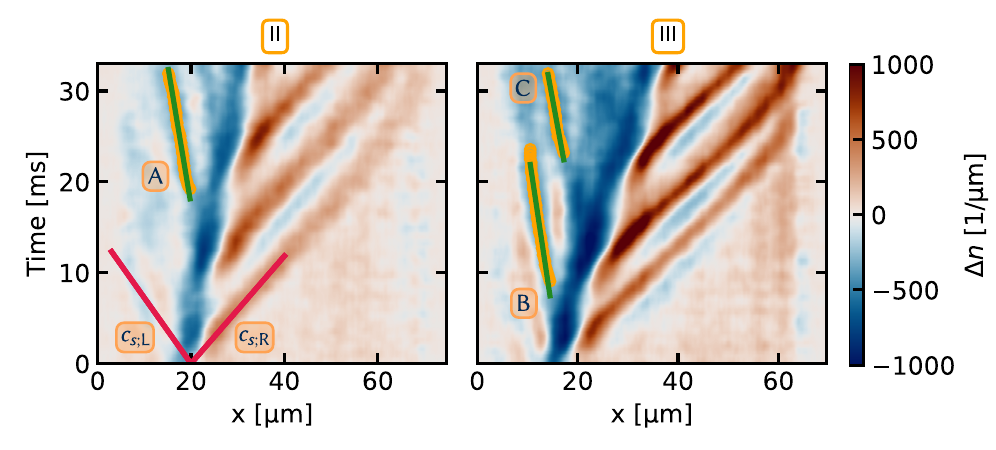}
\caption{
\textbf{Solitonic excitation velocity.} We extract the movement of the most prominent solitonic excitations and use a linear fit to estimate their velocities. 
The data are the ones from fig. 3 B II and III, additionally a Gaussian filter is applied. The solitonic exictations are labeled with $\{\mathrm{A,B,C}\}$. The red lines represent the speed of sound to the right/left side, derived in fig. S\ref{fig:speedofsound}. }
	\label{fig:soliton_velocity}
\end{figure}
}

\newcommand{\figvortexring
}{
\begin{figure}
\includegraphics[width=1\columnwidth]{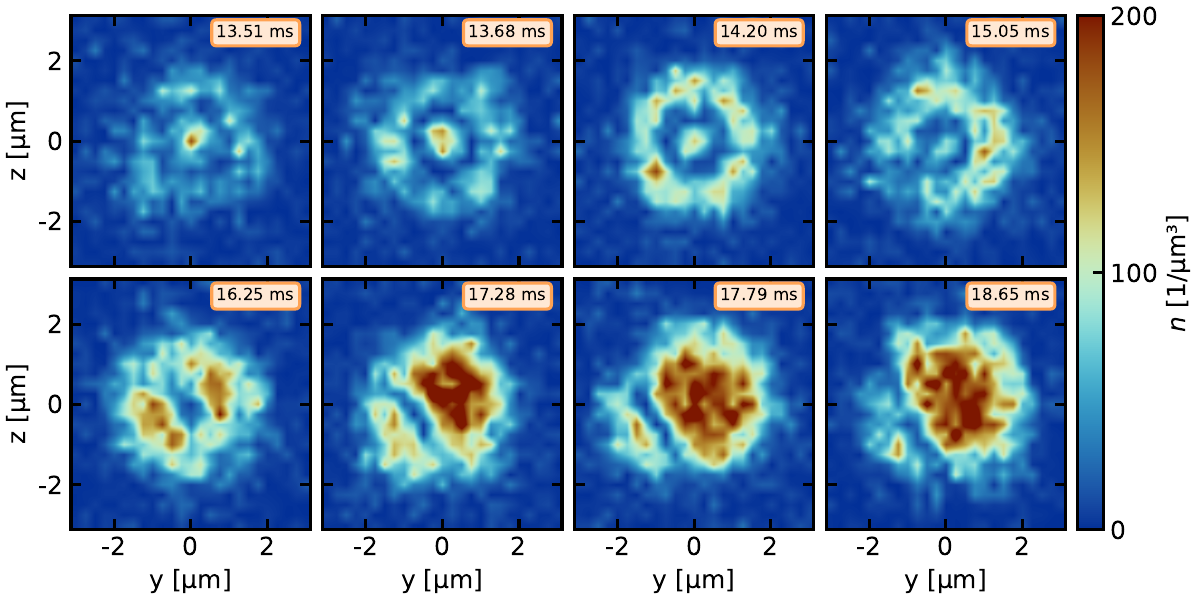}
\caption{
\textbf{Nucleation of a vortex ring.} Simulated transverse density profile of a single realization showing the solitonic excitation directly after its emergence. Initially, a vortex ring is created at the barrier, which decays into a vortex line within a few milliseconds. }
	\label{fig:vortex_ring}
\end{figure}
}

\subsection{Experimental Details}

%\subsection{Trap and barrier generation}

\subsubsection{Experimental procedure}

We start by preparing a Bose-Einstein condensate (BEC) of around $\SI{180e3}{}$ $^{87}\mathrm{Rb}$ atoms in a crossed optical dipole trap, consisting of a \textit{High Power} (HP) beam and a \textit{Low Power} (LP) beam. 
To reach the required tube like geometry we ramp the HP linearly in $\SI{400}{ms}$ from $\SI{20}{mW}$ up to $\SI{150}{mW}$ and the LP within the same time down to $ \SI{0}{mW}$.
This results in a trapping geometry with trapping frequencies $\omega = 2 \pi \times [1.6; 252; 250]$ Hz. 
Because the atoms can expand in the dipole trap during this procedure and to set a defined length of the sample we use two repulsive barriers at positions $x = \pm \SI{37.5}{\mu m}$ from the center of the initial position of the BEC.
The barriers are ramped linearly within $\SI{40}{ms}$ up to an intensity resulting in a potential much higher than the chemical potential of the cloud ($\approx 10 \mu$). 
      
When the ramp of the dipole trap is finished we wait $\SI{200}{ms}$, letting the system equilibrate, before starting the experiment.
After this wait time we ramp up the center barrier in $\SI{45}{ms}$ to its intended value with an initial position which is $\SI{20}{\mu m}$ off center. Subsequently, the Shapiro protocol starts, see main text.

\subsubsection{Barrier generation}

The realization of the optical barriers are sketched in the following.
We use a $\SI{532}{nm}$ laser to create a repulsive potential and an AOM to stabilize its intensity.
To generate the barriers we guide the beam through a two-axis acousto-optical deflector (AOD).
By changing the AODs RF driving frequency we control the axial position and the transverse extent of the barriers.
After the AOD we use a $f = \SI{100}{mm}$ scan lens to convert the beams angular displacement in a lateral displacement. 
The light is then collected by a $f = \SI{750}{mm}$ tube lens and guided to the $\mathrm{NA}=0.3$ in-vacuum objective, which focus the beam down to the atoms.     
The tube lens is chosen in a way that we get a diffraction limited spot at the atoms.
We estimate the size of a single spot by a measurement with a test target in the chamber. The Gaussian beam radius is around $\SI{1.1}{\mu m}$.
To realize in the transverse direction a homogeneously extended and in the axial direction movable barrier, we drive the AOD by multiple  RF frequencies \cite{EndresTweezer}. 
By applying a multi-tone RF signal of the form
\begin{equation}
S(t) = \sum_i S_i \sin(\omega_i t + \phi_i),
\end{equation}
to each of the AOD axis, we create three individual barriers, which are independently movable, when using time dependent $\phi_i(t)$. 
For the experiments described in the main text we use $\SI{40}{mW}$ laser power in total.

%\barriergeneration

\subsubsection{Calibration of the barrier height}

\chempotcalib

To estimate the height of the barrier used in the experiment we measure the equation of state $n(\mu)$. The method is close to the one used in \cite{Roati_acdc_jj_2020}. 
We generate a block of $5 \times 9$ barrier spots, which corresponds to 5 times the experiments barrier size and measure their mean intensity separately on a camera. 
This barrier block is projected onto the center of the BEC consisting of $\SI{180e3}{}$ atoms. We count the atoms at the position of the block by absorption imaging in the y-direction, applying the matter wave imaging scheme.
The result is shown in fig. S\ref{fig:chem_pot_calib}.
The initial decrease is linearly fitted and the zero crossing is set as the barrier height corresponding to the chemical potential $V_0 = \mu$.

\subsubsection{Imaging system}

The atom cloud is probed via absorption imaging, using the \SI{780}{nm} rubidium D2 transition. 
Our experiment uses three different imaging systems.
First we have a standard time of flight imaging along the y-axis (horizontal direction) with a magnification of $M_\mathrm{hor} = \SI{3.77}{}$, which is used to measure the atom number of the BEC.
Secondly, we use the same imaging setup in a different mode, known as matter wave imaging \cite{Asteria_2021}.
This procedure enables us to image the density distribution of the cloud in position space after a certain time of flight with a magnification in the axial direction of $M_\mathrm{mwi} = M_\mathrm{mwi}' \cdot M_\mathrm{hor} = \SI{32.8}{}$, were $M_\mathrm{mwi}' = \SI{8.7}{}$. 
The benefit of the matter wave imaging system is a strongly reduced optical density, which allows for a precise atom number determination.

Lastly, we can perform \textit{in situ} absorption imaging through the $\mathrm{NA} = \SI{0.3}{}$ objective along the z-axis (vertical direction).
With this imaging scheme we can get a theoretical optical resolution of $d_\mathrm{vert} = \SI{1.6}{\mu m}$ and a magnification of $M_\mathrm{vert} = \SI{19.5}{}$.
However, the high optical density of the sample in the trap prevents us from correctly determining the density with \textit{in situ} imaging.

\section{Characterization of the weak link} \label{weak_link}

\rcsj

To effectively describe a Josephson junction in the undriven case, one usually applies the so-called \textit{resistively capacitively shunted junction} (RCSJ) model. 
There, an additional resistance $R$ and capacity $C$ is assigned to the junction and a simple circuit, see fig. S\ref{fig:rcsj}, is used to quantitatively model the junctions behavior.
With Kirchoff's law we get the basic equation of the circuit
\begin{equation}
I_\mathrm{ext} = I_c \sin(\phi) -\frac{1}{R} \Delta \mu - C \Delta \dot{\mu}.
\end{equation}
Here, the first Josephson equation $I(\phi) = I_\mathrm{c} \sin(\phi)$ is used to describe the supercurrent across the junction.
In case of ultracold atoms, where the chemical potential difference $\Delta \mu$ plays the role of the voltage, the second Josephson equation reads \cite{Meier_Zwerger_2001}
\begin{equation}
\label{eq:josephson2}
\Delta \mu = - \hbar \dot{\phi}.
\end{equation} 
To be able to fully describe the junction we use these equations to experimentally determine the systems parameters $R$, $C$, and $I_\mathrm{c}$.

\figsystemcharc

To this end, we measure Josephson plasma oscillations \cite{Roati_dc_jj_scince_2015}. We first prepare an initial atom number imbalance across the junction and let the system evolve freely in time. 
We then measure both, the atom number difference $N = N_\mathrm{L} - N_\mathrm{R}$ and the relative phase $\phi = \phi_\mathrm{L} - \phi_\mathrm{R}$ between the two superfluids.   
Applying a Gaussian filter to the experimental data and numerical differentiation of $N$ gives $\frac{dN}{dt}$, and by fitting  $I_c$ to  $I(t) = \frac{dN}{dt} = I_\mathrm{c} \sin(\phi(t))$ (see fig. S\ref{fig:systemcharac}), we find $I_\mathrm{c} = \SI{{192\pm 7}e3}{1/s} $.
In the same way, we can determine the capacity $C$.
Using the same measurement, we numerically differentiate $\phi(t)$ and together with Eq. \ref{eq:josephson2} we get $ - V(t) = - \frac{N(t)}{C} = \frac{d\phi}{dt}$. 
We find $C = \SI{65.0 \pm 3.9}{s} \times 1/h$.  

\figmeasresistance

The resistance $R$ can be determined by measuring the BEC critical velocity $v_\mathrm{c}$. Thereby, we run a dc Josephson protocol, by moving the barrier with constant velocity through the condensate. 
The result is shown in fig. S\ref{fig:measresistance}. Below the critical velocity $v_\mathrm{c}$, we find $\Delta \mu = 0 $, indicating the absence of any ''voltage'' drop. Above $v_\mathrm{c}$, a finite chemical potential difference $\Delta \mu $ builds up, signaling the onset of the ac Josephson branch.
Fitting $\Delta z = R \sqrt{v^2 - v^2_\mathrm{c}}$, gives $v_\mathrm{c} = \SI{0.420 \pm 0.001}{mm/s}$. 
With $I = v \cdot I_\mathrm{c}/v_\mathrm{c}$ and Eq. \ref{eq:chem_pot} we can rescale the measurement data and use again the fit function $\Delta \mu = R \sqrt{I^2 - I^2_\mathrm{c}}$, and obtain $R=\SI{{907 \pm 4}e-6}{} \times h$.

\subsection{Current phase relation}

\currentphase

The Josephson junction in our experiments has a barrier height which is significantly lower than the chemical potential. It is necessary to independently measure the current phase relation in order to verify, whether higher order terms in the Josephson relation are relevant \cite{RevModPhys.76.411}.
We repeat the experiment for the critical velocity and measure the phase between the two condensates at the end of the motion.
The phase is determined via interference of both condensates in time of flight.
We find the best contrast for \SI{5}{ms} time of flight, for which a substantial part of the cloud has already interfered with each other.
After subtracting a reference image without barrier we fit a sinusoidal function to extract the phase.
The resulting current phase relation is shown in fig. S\ref{fig:cpr}.
Following Ref.\,\cite{Meier_Zwerger_2001} one can describe the current phase relation of a Josephson junction with $I(\varphi) = \sum_{n> 0} I_n \sin(n\varphi)$ if the system exhibits time reversal symmetry. 
In the limit of small coupling the above expression reduces to the first Josephson equation, $ I(\varphi) = I_\mathrm{c} \sin(\varphi)$.
We fit our data up to the second order term and find $I_1=\SI{0.995}{I_c}$ and $I_2=\SI{0.054}{I_c}$. The Josephson junction is therefore close to the ideal Josephson junction limit.

\subsection{Speed of sound}
\speedofsound

We measure the speed of sound following the original protocol from Ref.\,\cite{Andrews_1997}.
We prepare the same sample as in the Shapiro step measurements. We then instantaneously switch on a barrier with height $V_\mathrm{B} \approx \mu$ in the center of the condensate.
We follow the resulting phonon propagation via matter wave imaging, subtracting a reference image without barrier (see fig. S\ref{fig:speedofsound}). We extract from a linear fit the speed of sound. 
The density wave traveling to the right has a speed of $c_\mathrm{s} = \SI{1.63 \pm 0.03 }{mm/s}$, and the one traveling to the left has a speed of $c_\mathrm{s} = \SI{-1.38 \pm 0.02 }{mm/s}$. The small difference probably arises from the residual motion of the condensate in the shallow trap.

\subsection{Analysis of the solitonic excitations}
\figsolitonvelocity

The determination of the velocity of the solitonic excitations is depicted in fig. S\ref{fig:soliton_velocity}.
We apply a Gaussian filter to the experimental data of fig. 3 B II and III and extract the movement of the most prominent solitonic excitations labeled $\{A, B, C\}$. By using a linear fit we get their velocities, where we find $v_s/c_s = \{0.24,0.20,0.24\}$, with $c_s = \SI{-1.38}{mm/s}$, see above.

From the numeric simulations, we can also extract the transverse density profile of the solitonic excitation right after its emergence from the barrier. This is shown in fig. S\ref{fig:vortex_ring}. It is clearly visible that it is initially a vortex ring, which is emitted from the barrier and subsequently breaks. Due to the finite temperature, the contours are noisy and distorted. The continuing dynamics of decay will be the subject of future studies. For the Shapiro steps themselves, only the initial nucleation of  solitonic excitation at the barrier is of relevance.

\figvortexring

%\figsolitonresolution

\subsection{Shapiro steps for different barrier height}

To see the influence of the barrier height on the occurrence of the Shapiro steps, we repeat the same protocol for a large range of different barrier heights as shown in fig. S\ref{fig:stsh_barrier}.
Shapiro steps are visible over the whole parameter range. However, the steps tend to smear out for very high barriers (corresponding to a smaller critical current), while for the lowest barrier height ($V_\mathrm{B} =0.25 \mu$), the step height is reduced. We attribute the latter to the fact that the “weak link regime” is no longer valid for such low barriers.

\figstshbarrheight

\section{Measurement evaluation}

\subsection{Chemical potential}

\chempot

To map the experimentally measured atom number differences $\Delta z$, see main text, to a chemical potential difference $\Delta \mu$, we numerically calculate $\mu$ for different total atom numbers $N$, see fig. S\ref{fig:chem_pot}.
Therefore, we use the measured experimental parameters, as the trapping frequency and the geometry described in the section 'Experimental procedure', to model the trapping potential, without the barrier. This potential is put in the Gross-Pitaevskii equation to calculate the ground-state wavefunction, using an imaginary time evolution of the GPE \cite{BaalsStability}.
We then extract the chemical potential from the calculated ground state wave function, and use a phenomenological fit to get a function $\mu (N)$, where we get $\mu (N) = \SI{4.619}{Hz} \cdot N^{1/2}$.
With \cite{Giovanazzi_2000,singh2023shapiro}
\begin{equation}
\label{eq:chem_pot}
\Delta \mu (\Delta z,N) = \mu(N\cdot(1+ \Delta z)) - \mu(N\cdot(1- \Delta z))
\end{equation}
we calculate $\Delta \mu(\Delta z)$.

\subsubsection{Step height}

To evaluate the step height, we rescale the data to chemical potential differences using Eq. \ref{eq:chem_pot} and fit a logistic function 
\begin{equation}
f(x) = \sum_i^M \frac{S}{1+\exp(-k_i(x-x_i))} + b,
\end{equation}
where $S$ is the step height, $k_i$ the steepness of the steps and $M-1$ the number of visible steps.  
%The displayed values in \ref{fig:freqdep} are the mean of $S$ for the unique frequencies.

\subsubsection{Step width}

We extract the width of the steps as illustrated in fig. S\ref{fig:stepwidth_single}.  
We first apply a Gaussian filter to the data to reduce high frequency noise before we numerically calculate the first derivative $\frac{\partial \Delta z}{\partial v}$ of the recorded atom number imbalance $\Delta z$. 
For each step $n$ we search the maxima of the derivative in the area where $\Delta \mu = n f_\mathrm{m} \pm \epsilon$, with $\epsilon = \SI{45}{Hz}$.  
We define the width of a step $I_n$ as the distance between the two maxima. 
For the $0$-th step we calculate the distance with respect to zero.
$I_n$ is normalized by $I_0$, which is the width of the $0$-th step in the dc Josephson protocol.

\figstepwidthsingle

%%%--%%%5--Simulation method----%%%----%%%%
%
\section{Classical-field simulation method}
We simulate the dynamics of a driven atomic Josephson junction using a classical-field method within the truncated Wigner approximation \cite{Singh2016, Kiehn2022}. 
We consider a three-dimensional (3D) condensate of  $^{87}$Rb atoms confined in a cigar-shaped geometry. 
The system is described by the Hamiltonian
\begin{align} \label{eq:hamil}
\hat{H} &= \int \mathrm{d}{\bf r} \Big[  \frac{\hbar^2}{2m}  \nabla \hat{\psi}^\dagger({\bf r}) \cdot \nabla \hat{\psi}({\bf r}) + V({\bf r}) \hat{\psi}^\dagger({\bf r})\hat{\psi}({\bf r})
   \, \nonumber \\ \quad &+ \frac{g}{2} \hat{\psi}^\dagger({\bf r})\hat{\psi}^\dagger({\bf r})\hat{\psi}({\bf r})\hat{\psi}({\bf r})\Big],
\end{align}
where $\hat{\psi}({\bf r})$ and $\hat{\psi}^\dagger({\bf r})$ are the bosonic annihilation and creation field operators, respectively. 
The 3D interaction parameter is given by $g=4\pi a_s \hbar^2/m$, where $a_s$ is the $s$-wave scattering length and $m$ is the mass. 
For $^{87}$Rb atoms $a_s$ is $5.3\, \mathrm{nm}$.  
The external potential $V({\bf r})$ represents the harmonic trap $V_{\mathrm{trap}}({\bf r})=m(\omega_x^2x^2+ \omega_y^2 y^2 + \omega_z^2 z^2)/2$,  where the trap frequencies are chosen according to the experiment, i.e.,  $(\omega_x, \omega_y, \omega_z)= 2\pi \times(1.6, 252, 250)\, \mHz$.  
Within the classical-field approximation we replace the operators $\hat{\psi}$ in Eq. \ref{eq:hamil} and in the equations of motion by complex numbers $\psi$.  We map real space on a lattice system of  $508 \times 33 \times 33$ sites with the lattice discretization length $l= 0.15\, \mu \mm$. 
We note that the continuum limit is satisfied by choosing $l$ to be comparable or smaller than the healing length $\xi = \hbar/\sqrt{2mgn}$ and the de Broglie wavelength, where $n$ is the density \cite{Mora2003}. 
We generate the initial states $\psi(t=0)$ in a grand canonical ensemble of temperature $T$ and chemical potential $\mu$ via a classical Metropolis algorithm. We use $T= 35\, \mnK$  and adjust $\mu$ such that the total atom number $N$ is close to the experimental one. 
We propagate each initial state using the classical equations of motion. 

To create a Josephson junction and excite Shapiro steps we add a perturbation term $\mathcal{H}_{ex} = \int \mathrm{d}{\bf r}\, V(x,t) n({\bf r}, t)$, where $n({\bf r}, t) = |\psi({\bf r} , t)|^2$ is the local density and $V(x, t)$ is the Gaussian barrier potential of the form 
\begin{equation}\label{eq:pot} 
V(x,t)  = V_0 (t) \exp \Bigl[- \frac{ 2\bigl( x-x_0- x(t) \bigr)^2}{w^2} \Bigr].
\end{equation}
$V_0$ is the time-dependent strength and $w$ is the width. $x_0$ is the initial location of the barrier and $x(t)$ is the dc and ac driving term. 
We choose $w=1 \, \mum$ and $x_0= 19\, \mum$. 
We linearly ramp up $V_0(t)$ to the value $V_0/\mu=0.45$ over $200\, \mms$ and then wait for $50\, \mms$.
This creates a weak link by suppressing the tunneling at location $x_0$. 
While we move the barrier at a constant velocity $v$,  we also periodically modulate its position as 
\begin{equation}
x(t) = v t + x_\mm \sin(2\pi f_\mm t),
\end{equation}

where $x_\mm$ is the driving amplitude and $f_\mm$ is the driving frequency \cite{singh2023shapiro}.   
This induces an atom current $I$ relative to the barrier motion, i.e., $I = v I_c/v_c$, 
where $I_c$ is the critical current and $v_c$ is the critical velocity. 
Similarly, the amplitude of the ac current is given by $I_\mm= 2\pi f_\mm x_\mm I_c/v_c$. 
For the calculation of the atom imbalance we fix the driving time to $33\, \mms$. 
The atom imbalance  $z$ is determined by $z=(N_R - N_L)/N$, 
where $N_L$ ($N_R$) is the atom number in the left (right) reservoir, and $N= N_L +N_R$ is the total atom number.  

\figsimfreq

\figsimsoliton

For various values of $I$, $I_\mm$, and $f_\mm$ we determine $\Delta z = z - z_\mref$ and convert it to the chemical potential difference $\Delta \mu$ using Eq. \ref{eq:chem_pot}. 
$ z_\mref$ is the imbalance determined at the final location without the barrier. 
Without the ac driving the onset of a nonzero $\Delta \mu$ occurs above a certain $v_c$, which we determine using the prediction of the RCSJ circuit model, see Fig. S\ref{fig:simfreq} A. We obtain $v_c=0.42\, \mathrm{mm/s}$ in excellent agreement with the measurement of $v_c$. 
In the presence of ac drive we find the creation of Shapiro steps whose height depends on the driving frequency $f_\mm$. Using sigmoid fits we determine the height of the first Shapiro step, see Fig. S\ref{fig:simfreq} B. 
We determine the step height for $f_\mm$ in the range between $50$ and $185\, \mHz$ by analyzing the driven response after $3$ to $6$ driving periods, and also average these results over a few values of $I_\mm$. 
We determine the width of zeroth and first step following the same procedure as in the experiment, 
which involves calculating the differential resistance $d \Delta \mu/dI$ and 
then determining the step width from the maximum of $d \Delta \mu /dI$, see Fig. S\ref{fig:stepwidth_single}. 
The simulation results of the step width for $f_\mm=90\, \mHz$ and varying $I_\mm/I_c$ are 
presented in the main text of the paper. 
For the phase evolution we calculate the local phase $\delta \phi ({\bf r})=  \phi ({\bf r}) -  \phi({\bf r}_\mref)$ from the time evolution of a single trajectory, 
where $\phi({\bf r}_\mref)$ is the reference phase.

To count total number of solitonic excitations we analyze the phase evolution $\delta \phi (x, t)=  \phi (x+l, t) -  \phi(x, t)$ of a single line along the $x$ direction of a single trajectory of the driven system, 
which is because the phase profile in the $yz$ plane is initially almost uniform. 
The oscillating motion of the barrier results in the creation of solitonic excitations, which we identify by a phase jump of less than $\pi$. 
We count total number of such phase jumps to determine the solitonic excitation number $N_s$. 
In Fig. S\ref{fig:simsoliton}B we show $N_s$, averaged over few samples and total number of driving cycles, as a function of $I/I_c$, which features a step-like behavior, coinciding with the one observed in the $I- \Delta \mu$ characteristic in Fig. S\ref{fig:simsoliton}A. There are on-average one, two and three solitonic excitation at the first, second and third Shapiro steps, respectively.

\end{document}